\documentstyle[aps,prl]{revtex} 
\begin{document}
%\clubpenalty 10000
%\widowpenalty 10000

\draft
\twocolumn
\narrowtext
\title{Unconditionally secure quantum bit commitment is impossible}
\author{Dominic Mayers}
\address{D\'epartement IRO, Universit\'e de Montr\'eal \\
C.P. 6128, succursale Centre-Ville,Montr\'eal (Qu\'ebec), Canada H3C 3J7.}
\date{\today}
\maketitle

\begin{abstract}
The claim of quantum cryptography has always been that it can provide
protocols that are unconditionally secure, that is, for which the
security does not depend on any restriction on the time, space or
technology available to the cheaters.  We show that this claim does
not hold for any quantum bit commitment protocol.  Since many
cryptographic tasks use bit commitment as a basic primitive, this
result implies a severe setback for quantum cryptography. 
The model used encompasses all reasonable implementations of
quantum bit commitment protocols in which the participants have not
met before, including those that make use of the theory of special
relativity.
\end{abstract}


\paragraph{Introduction.}
Quantum cryptography is often associated with a cryptographic
application called key distribution~\cite{bb84,bbbss90} and it has
achieved success in this area~\cite{bc96}.  However, other
applications of quantum mechanics to cryptography have also been
considered and a basic cryptographic primitive called bit commitment,
the main focus of this letter, was at the basis of most if not all of
these other applications~\cite{bbcs92,bcjl93,yao95,bc96}.

%For instance, the work of Kilian and many
%others~\cite{kilian88,gmw87,cdg87,cgt95} has shown that
%any multi-party oblivious function
%evaluation can be reduced to two-party oblivious 
%transfers~\cite{bbcs92} which
%in turn can be implemented by a quantum protocol supported by some means
%of secure bit commitment. 
%In a multi-party computation each participant has an input $x_i$
%and receives an output $y_i(x_1, \ldots, x_n)$ and as little additional
%information about these inputs as possible. An
%example of multi-party computation is the {\em millionaire's problem}.
%In this task, two millionaires want to find out who is the richer,
%but they do not want to unveil any additional information about their
%respective fortune.  

In a concrete example of bit commitment, a party, Alice, writes a bit
$b$ on a piece of paper and puts it into a safe.  She gives the safe
to another party, Bob, but keeps the key.  The objective of this
scheme, and of bit commitment in general, is that Alice cannot change
her mind about the value of the bit $b$, but meanwhile Bob cannot
determine the bit $b$.  At a later time, if Alice wants to unveil $b$
to Bob, she gives the key to Bob.

In 1993, a protocol was proposed to realize bit commitment in the
framework of quantum mechanics, and the unconditional security (see
sections~\ref{model} and \ref{definition_security}) of this
protocol has been generally accepted for quite some time. However,
this result turned out to be wrong.  The non security of this
protocol, called the BCJL protocol, was realized in the fall of
1995~\cite{mayers95b}.  After this discovery, Brassard, Cr\'epeau and
other researchers have tried to find alternative
protocols~\cite{brassard96}.  Some protocols were based on the theory
of special relativity.  For additional information about the history
of the result see~\cite{bc96}.  See also~\cite{lc96a}.

Here it is shown that an unconditionally secure bit commitment
protocol is impossible, unless a computing device, such as a beam
splitter, a quantum gate, etc.\ can be simultaneously trusted by both
participants in the protocol.  This encompasses any protocol based on
the theory of special relativity.  A preliminary version of the proof
appeared in~\cite{mayers96b}.

\paragraph{The model for quantum protocols.} \label{model}
It is neither possible in this letter to describe in detail a model
for two-party quantum protocols, nor is it is useful for the purpose
of this letter. The following description includes all that is
necessary for our proof.

In our model, a two-party quantum protocol is executed on a system
$H_A \otimes H_B \otimes H_E$ where $H_A$ and $H_B$ correspond to two
areas, one on Alice's side and one on Bob's side, and $H_E$
corresponds to the environment.  We adopt the ``decoherence'' point of
view in which a mixed state $\rho$ of $H_A \otimes H_B$ is really the
reduced state of $H_A \otimes H_B$ entangled with the environment
$H_E$, the total system $H_A \otimes H_B \otimes H_E$ always being in
a pure state $|\psi\rangle$.  The systems $H_A$ and $H_B$ contain only
two dimensional quantum registers.  Higher dimensional systems can be
constructed out of two dimensional systems.  Alice and Bob can execute
any unitary transformation on their respective system.  In particular,
they can introduce new quantum registers in a fixed state $|0\rangle$.
%The number
%of quantum registers in $H_A \otimes H_B$ is not fixed.  
States that correspond to different number of registers can be in
linear superposition.  Any mode of quantum communication can be
adopted between Alice and Bob.

Without loss of generality, we can restrict ourselves to binary
outcome measurements.  The environment is of the form $H_E = H_S
\otimes H_{E,A} \otimes H_{E,B}$ where $H_S = H_{S,A} \otimes H_{S,B}$
is a system that stores classical bits that have been transmitted from
$H_{S,A}$ on Alice's side to $H_{S,B}$ on Bob's side or vice versa,
and $H_{E,A}$ and $H_{E,B}$ store untransmitted classical bits that
are kept on Alice's side and Bob's side respectively.  To execute a
binary outcome measurement, a participant $P \in \{A,B\}$, where $A$
and $B$ stand for Alice and Bob respectively, introduces a quantum
register in a fixed state $|0\rangle$.  The participant $P$ entangles
this register with the measured system initially in a state
$|\phi\rangle$ and obtains a new state of the form $\alpha\, |0\rangle
|\phi_0\rangle +\beta\, |1\rangle |\phi_1\rangle$.  Then, he sends the
new quantum register away to a measuring apparatus in $H_{E,P}$ which
amplifies and stores each component $|x\rangle$ as a complex state
$|x\rangle^{(E,P)}$.  The resulting state is $\alpha \,
|0\rangle^{(E,P)} |\phi_0\rangle + \beta \, |1\rangle^{(E,P)}
|\phi_1\rangle$.
%corresponds to the binary
%outcome of the measurement with
%probability distribution $p_0 = |\alpha|^2$ and $p_1 = |\beta|^2$.
Similarly, to generate a random bit one simply maps $|0\rangle$ into
$\alpha \, |0\rangle + \beta \, |1\rangle$ and sends the register away
in some part of $H_{E,P}$ that will amplify and store it as a state
$\alpha \, |0\rangle^{(E,P)} + \beta \, |1\rangle^{(E,P)}$.  The
transmission of a classical bit $x$ from Alice to Bob is represented
by a transformation that maps $|x\rangle^{(E,A)}|0\rangle^{(E,B)}$
into $|x\rangle^{(S,A)}|x\rangle^{(S,B)}$. A similar transformation
exists for the transmission of a classical bit from Bob to Alice.

Now, let us assume that the total system is in a superposition
$\sum_{\xi_S,\xi_A,\xi_B} \alpha_{(\xi_S,\xi_A,\xi_B)} \,
|\xi_S,\xi_A,\xi_B\rangle^{(E)} |\phi_{(\xi_S,\xi_A,\xi_B)}\rangle$
where $|\xi_S,\xi_A,\xi_B\rangle^{(E)}$ corresponds to the random
binary string stored in the environment with probability
$|\alpha_{(\xi_S,\xi_A,\xi_B)}|^2$ and
$|\phi_{(\xi_S,\xi_A,\xi_B)}\rangle$ is the corresponding collapsed
state of $H_A \otimes H_B$.  The participant $P$ can ``read'' the
strings $\xi_P$ and $\xi_S$ and then choose the next action,
measurements, etc.\ accordingly, but the allowed transformations must
behave as if a collapse into the state
$|\phi_{(\xi_S,\xi_A,\xi_B)}\rangle$ has really occurred.

\paragraph{Unconditional security and quantum bit commitment protocols.}
\label{definition_security}
To realize bit commitment in the framework of quantum mechanics, the
bit $b$ that Alice has in mind must be encoded into a state
$|\psi_b\rangle$ of $H_A \otimes H_B \otimes H_E$ through a procedure
$commit(b)$.  A bit commitment protocol must also include an optional
procedure $unveil(|\psi_b\rangle)$ that can be used to return to Bob
either the value of the bit $b$ or, occasionally when Alice attempts
to cheat, an inconclusive result denoted $\perp$.  The protocol is
{\em correct} if the procedure $unveil$ always return $b$ on
$|\psi_b\rangle$ when both participants are honest.

Now, the encoding that is defined above does not always make sense
when Alice cheats. Alice might act without having any specific bit $b$
in mind during the procedure $commit$, so as to choose it later.
Given a fixed strategy used by Alice, let $|\psi'\rangle$ be the state
created by the associated modified procedure $commit'$.  We denote
$p(b \, | \mbox{not} \perp)$ the probability that $unveil$ returns $b$
on $|\psi'\rangle$ given that it has not returned $\perp$.  Alice can
certainly choose the probability $p(b\, | \mbox{ not} \perp)$.  This
can be done via an honest encoding by choosing bit $b$ with
probability $p( b \,| \mbox{ not} \perp)$.  However, after the
procedure $commit'$, Alice should not be able to change her mind about
$p(b\, | \mbox{ not} \perp)$.  Let $unveil'$ be a procedure $unveil$
modified by a dishonest Alice.  Now, denote $p'(b \, | \mbox{ not}
\perp)$ as the probability that $unveil'$ returns $b$ on
$|\psi'\rangle$ given that it does not return $\perp$.  The state
$|\psi'\rangle$ {\em perfectly binds} Alice to $p(b\, | \mbox{ not}
\perp)$ if every procedure $unveil'$ either returns $\perp$ with
probability $1$ or else returns $b$ with probability $p'(b\, | \mbox{
not} \perp) = p(b\, | \mbox{ not} \perp)$.  In this case, we also say
that $|\psi'\rangle$ is {\em perfectly binding}.

The encoding $b \mapsto |\psi_b\rangle$ makes sense when Alice is
honest, but it can be modified by a dishonest Bob.  Let $\eta =
(\xi_B,\xi_S)$ be the random classical information stored in $H_{E,B}
\otimes H_S$ and available to Bob after this encoding. Let
$|\psi_{b,\eta}\rangle$ be the corresponding collapsed state of the
system $H_A \otimes H_B \otimes H_{E,A}$.  Denote
$\rho_B(|\psi_{b,\eta} \rangle) = {\rm Tr}_{H_A \otimes
H_{E,A}}(|\psi_{b,\eta}\rangle \langle \psi_{b,\eta}|)$ the reduced
density matrix of $H_B$ given $\eta$.  Let us define $F(\eta) = 0$ if
$\eta$ determines a single value of the bit $b$, otherwise let
$F(\eta)$ be the fidelity~\cite{jozsa94} between
$\rho_B(|\psi_{0,\eta} \rangle)$ and $\rho_B(|\psi_{1,\eta} \rangle)$.
The fidelity is never greater than $1$ and is equal to $1$ if and only
if the two density matrices are identical.  The modified encoding is
said to be {\em perfectly concealing} if the random string $\eta$
provides no information about $b$ and the expected value of $F(\eta)$
is $1$.  This corresponds to the fact that a dishonest Bob should not
be able to determine the bit $b$.  A protocol is {\em perfectly
secure} if, (1) when Alice is honest, even if Bob cheats, the
resulting encoding is perfectly concealing, and, (2) when Bob is
honest, even if Alice cheats, the resulting encoding is perfectly
binding.

Note that it is generally accepted that a perfectly secure bit
commitment protocol is impossible.  However, another almost as
interesting level of security is possible.  Consider a protocol with
some security parameter $n$.  For example, the security parameter $n$
could correspond to the number of photons that must be transmitted.
An encoding with parameter $n$ is said to be {\em concealing} if, by
an increase of the parameter $n$, it can be made arbitrarily close to
perfectly concealing.  Similarly, a state $|\psi\rangle$ with an
implicit parameter $n$ is said to be {\em binding} if by an increase
of the parameter $n$ it can be made arbitrarily close to be perfectly
binding.  A protocol with parameter $n$ is {\em secure} if (1) the
state $|\psi\rangle$ returned by $commit$ is binding when Bob is
honest and (2) the encoding is concealing when Alice is honest. This
is the kind of security that we expect in quantum cryptography.
Furthermore, in quantum cryptography, we want any desired properties
to hold even against a cheater with unlimited computational power!
This means that there should be no restriction on the amount of time,
space or technology available to the cheater.  A property that holds
even against such a cheater is said to hold {\em unconditionally}.  In
quantum cryptography, we want unconditionally secure protocols.  This
does not mean that we want perfectly secure protocols.

%All the known protocols that attempt to realize bit commitment via
%exchange of classical informations only are based upon a restriction
%on the amount of time available to the cheater.  This is what we call
%computational security.  Furthermore, they are all based
%on some unproven computational assumptions such as the assumption that
%no algorithm can factor a large number in a reasonable number of
%steps.

\paragraph{The BB84 quantum bit commitment protocol.}
We say that an encoding $b \mapsto |\psi_b\rangle$ is a {\em bit
commitment encoding} if it is concealing and $|\psi_0\rangle$ and
$|\psi_1\rangle$ bind Alice to $0$ and $1$ respectively.  It can be
shown that even if both participants are honest, no protocol that is
based on classical communication between Alice and Bob can create a
bit commitment encoding.  So, it is of interest that a two-party
quantum protocol was proposed in 1984 that realizes a bit commitment
encoding when both participants are honest~\cite{bb84}.  The protocol
fails when Alice cheats.  In fact, the authors themselves have first
explained their protocol together with Alice's strategy.

In the BB84 coding scheme (which is not a bit commitment) a bit is
coded either in a so-called rectilinear basis $(\,|0\rangle_+,\,
|1\rangle_+ \,)$ or in the diagonal basis $(\,|0\rangle_\times,\,
|1\rangle_\times \,)$, where $|0\rangle_\times = 1/\sqrt{2} (
|0\rangle_+ + |1\rangle_+)$ and $|1\rangle_\times = 1/\sqrt{2} (
|0\rangle_+ - |1\rangle_+)$.  In the $commit$ procedure of the BB84
quantum bit commitment protocol, Alice creates a string of random bits
$w = w_1 \ldots w_n$.  Then she codes each bit $w_i$ in the BB84
coding scheme, always using the rectilinear basis $\theta = +$ if she
wants to commit a $0$ and the diagonal basis $\theta = \times$ if she
wants to commit a $1$.  She sends these registers to Bob.  Then, Bob
chooses a string of random bases $\hat{\theta} = \hat{\theta}_1 \ldots
\hat{\theta}_n \in \{+,\times\}^n$, measures the register $i$ in the
basis $\hat{\theta}_i$ and notes the outcomes $\hat{w}_i$.  In the
$unveil$ procedure, Alice has simply to announce the string $w$.  Bob
can determine the bit $b$ by looking at the positions $i$ where $w_i
\neq \hat{w}_i$.  Bob knows that at each of these positions $\theta
\neq \hat{\theta}_i$, and he knows the bases $\hat{\theta}_i$.  Any of
these positions can be used to determine $\theta$.  If two of these
positions reveal different values for $\theta$, Bob interprets it as
an inconclusive result.  The encoding is concealing because both $b =
0$ and $b = 1$ correspond to the same fully mixed density matrix on
Bob's side.  Also, the state after the $commit$ procedure is binding
because in order to deceive Bob Alice would have to guess exactly the
bits obtained by Bob when $\hat{\theta}_i \neq \theta$. These bits are
perfectly random.  Therefore, she would only succeed with a
probability that goes to $0$ when $n$ increases.  Note that
unconditional security does not mean a perfectly secure protocol.

Now, we present Alice's strategy against the BB84 bit commitment
protocol. In our model, for each random bit $w_i$, Alice creates the
state:
\begin{equation} \label{honest_bb84_0}
1/\sqrt{2}( |0\rangle_\theta^{(E,A)} 
|0\rangle_\theta^{(B)} + |1\rangle_\theta^{(E,A)}
|1\rangle_\theta^{(B)} )
\end{equation}
where the bit $w_i$ is coded in the register to the left.  For
simplicity, we have assumed that the basis $\theta$ is used for both
registers. A dishonest Alice executes the honest commit algorithm for
$b = 0$, except that she never sends anything away to the environment.
In other words, for each position $i$, the state (\ref{honest_bb84_0})
becomes the state:
\begin{equation} \label{dishonest_bb84_0}
1/\sqrt{2}( |0\rangle_+^{(A)} |0\rangle_+^{(B)} +
|1\rangle_+^{(A)} |1\rangle_+^{(B)} ).
\end{equation}
Note that the states (\ref{honest_bb84_0}) and
(\ref{dishonest_bb84_0}) are formally identical. Only the underlying
systems are different.  Nevertheless, this is cheating because now
there exists a unitary transformation that Alice can execute on $H_A$
that will transform this state into the state:
\begin{equation} \label{dishonest_bb84_1}
1/\sqrt{2}( |0\rangle_\times^{(A)} 
|0\rangle_\times^{(B)} + |1\rangle_\times^{(A)}
|1\rangle_\times^{(B)} ),
\end{equation}
which is the state that she would have created with a $1$ in mind.  In
this example, it turns out that the transformation is the identity
transformation because these two states are one and the same state,
but in general the cheater will have a non trivial transformation to
execute.

\paragraph{The proof.}
It is very easy to build a secure bit commitment protocol in which the
initial state is already the outcome of a bit commitment encoding.  So
the following proof for the impossibility of bit commitment requires
an assumption on the initial state.  For simplicity we deal only with
protocols where initially all quantum registers are set to $|0\rangle$
and there is no entanglement with the environment.  We prove that no
quantum bit commitment protocol that starts in this state is
unconditionally secure, unless a computing device such as a beam
splitter can be trusted by both participants simultaneously.  In our
proof we assume that the protocol is secure against Bob.  (Otherwise,
the protocol is not secure and we are done).  The proof has three main
steps.  First, we describe Alice's strategy in a modified procedure
$commit'$ and Bob's strategy in a modified procedure $commit''$.
Second, we consider Bob's strategy in $commit''$ and use the
assumption that the protocol is secure against Bob to obtain that the
expected value of the fidelity between the density matrices on Bob's
side after $commit'$ is arbitrarily close to $1$.  Third, we show that
this implies that a procedure $unveil'$ modified by Alice allows her
to cheat after $commit'$.
 
{\em The first step.}  In the BB84 example, Alice's strategy in a
procedure $commit'$ was to choose $b = 0$ and to never send a register
away to the environment.  However, in this particular example there
was no classical communication from Alice to Bob.  In the general
case, in the modified procedure $commit'$, Alice chooses $b = 0$ and
never sends a register away to the environment except when this
register contains a classical bit that she must transmit to Bob via
the environment, using the phone for instance.  Bob in $commit''$ does
as Alice in $commit'$, that is, he never sends a register away to the
environment unless it is required for classical communication.  So,
$H_{E,A}$ is not used in $commit'$ and $H_{E,B}$ not used in
$commit''$.

{\em The second step.}  Let $\gamma$ be the random string stored in
$H_S$ after $commit'$.  Let $|\psi'_{b,\gamma}\rangle$ be the
corresponding collapsed state of the remaining system $H_A \otimes H_B
\otimes H_{E,B}$. We want to show that the expected value of the
fidelity $F'(\gamma)$ between the reduced density matrices
$\rho_B(|\psi'_{b,\gamma}\rangle)$ for $H_B \otimes H_{E,B}$ in
$commit'$ is arbitrarily close to $1$.  After $commit''$, the same
random string $\gamma$ is stored in $H_S$, but the corresponding
collapsed state $|\psi''_{b,\gamma}\rangle$ is now stored in $H_{E,A}
\otimes H_A \otimes H_B$.  However, as for the states
(\ref{honest_bb84_0}) and (\ref{dishonest_bb84_0}) of the BB84
example, the state $|\psi''_{b,\gamma}\rangle$ is formally identical
to the state $|\psi'_{b,\gamma}\rangle$.  Also, because in $commit'$
$H_{E,A}$ has been replaced by a subsystem of $H_A$, a partial trace
over $H_A$ in $commit'$ corresponds formally to a partial trace over
$H_A \otimes H_{E,A}$ in $commit''$. Therefore, the density matrices
$\rho_B(|\psi'_{b,\gamma}\rangle)$ in $commit'$ are identical to the
corresponding density matrices $\rho_B(|\psi''_{b,\gamma}\rangle)$ for
the system $H_B$ in $commit''$.  Also, in $commit''$ the strings $\eta
= (\xi_B,\xi_S)$ and the string $\gamma = \xi_S$ correspond to a same
collapse because $\xi_B$ is the empty string.  The expected value of
$F'(\gamma) = F(\eta)$ (see section \ref{definition_security}) must be
arbitrarily close to $1$, otherwise Bob succeeds in $commit''$ and
this contradicts our assumption.
  
{\em The third step.}  For simplicity we first do the case where the
expected value of $F'(\gamma)$ is $1$, that is, the density matrices
are always identical.  In this case, Alice can unveil the bit $b = 1$
because the work of \cite{hjw93} implies that, if $
\rho_B(|\psi'_{0,\gamma}\rangle) = \rho_B(|\psi'_{1,\gamma} \rangle)
\stackrel{def}{=} \rho_B , $ there exists a unitary transformation on
Alice's side which maps $|\psi'_{0,\gamma}\rangle$ into
$|\psi'_{1,\gamma}\rangle$.  Consider the respective Schmidt
decomposition \cite{hjw93,schmidt06} of $|\psi'_{0,\gamma}\rangle$ and
$|\psi'_{1,\gamma}\rangle$:
\begin{eqnarray*}
|\psi'_{0,\gamma} \rangle & = & \sum_i \sqrt{\lambda_i} |e^{(0)}_i\rangle
\otimes |f_i\rangle \\
|\psi'_{1,\gamma} \rangle & = & \sum_i \sqrt{\lambda_i} |e^{(1)}_i\rangle
\otimes |f_i\rangle.
\end{eqnarray*}
In the above formula, $\lambda_i$ are eigenvalues of the three density
matrices $\rho_B$, $\rho_A(|\psi'_{0,\gamma}\rangle)$ and
$\rho_A(|\psi'_{1,\gamma}\rangle)$.  The fact that these three density
matrices share the same positive eigenvalues with the same
multiplicity is a direct consequence of the Schmidt decomposition
theorem \cite{hjw93,schmidt06}.  The states $|e^{(b)}_i\rangle$ and
$|f_i\rangle$ are respectively eigenstates of
$\rho_A(|\psi'_{b,\gamma}\rangle)$ and $\rho_B$ associated with the
same eigenvalue $\lambda_i$.  The coefficients $\sqrt{\lambda_i}$ are
real numbers, but any phase can be included in the choice of
$|e^{(b)}_i\rangle$.  Clearly, the same unitary transformation that
maps $|e^{(0)}_i\rangle$ into $|e^{(1)}_i\rangle$ also maps
$|\psi'_{0,\gamma}\rangle$ into $|\psi'_{1,\gamma}\rangle$.  Alice can
compute the states $e^{(b)}_i$ and thus this unitary transformation
with an arbitrary level of precision.  So, Alice can cheat when the
two density matrices on Bob's side are always identical.

Now, we do the case where the expected value of $F'(\gamma)$ is not
$1$ but arbitrarily close to $1$. Note that $F'(\gamma) > 0$ is the
fidelity between $\rho_B(|\psi'_{0,\gamma}\rangle)$ and
$\rho_B(|\psi'_{1,\gamma}\rangle)$.  Any state $|\psi_{01}\rangle$ of
the overall system such that $\rho_B(|\psi_{01}\rangle) =
\rho_B(|\psi'_{0,\gamma}\rangle)$ is called a purification of the
density matrix $\rho_B(|\psi'_{0,\gamma}\rangle)$. Because
$|\psi'_{1,\gamma}\rangle$ is a purification of
$\rho_B(|\psi'_{1,\gamma}\rangle)$, Uhlmann's theorem~\cite{jozsa94}
says that there exists a purification $|\psi_{01}\rangle$ of
$\rho_B(|\psi'_{0,\gamma}\rangle)$ such that
\begin{equation} \label{Uhlmann_inequality}
\langle \psi_{01} | \psi'_{1,\gamma} \rangle \geq F'(\gamma)
\end{equation}  
The fact that $|\psi_{01}\rangle$ is a purification of
$\rho_B(|\psi'_{0,\gamma}\rangle)$ implies that Alice in $unveil'$ can
transform $|\psi'_{0,\gamma}\rangle$ into $|\psi_{01}\rangle$, as in
the case where the density matrices are identical, and then continue
with the honest $unveil$.  Inequality \ref{Uhlmann_inequality} implies
that the probability $p_\gamma$ that $unveil'$ returns $1$ on
$|\psi'_{0,\gamma}\rangle$ is greater than $f(F'(\gamma))$ for some
function $f(z)$ such that $\lim_{z \rightarrow 1} f(z) = 1$ (more
details are given in \cite{mayers95b}). This means that Alice can
change the bit $b$ that she unveils to Bob from $0$ to $1$ with a
probability that goes to $1$ as the expected value of $F'(\gamma)$
goes to $1$.

One key point is that the algorithm used by the dishonest participant
in $commit'$ or $commit''$ is formally identical to the algorithm used
by the same but honest participant in $commit$.  Therefore, no
verification whatsoever, including any verification based on
measurement of time delay and the theory of special relativity, can be
used by the honest participant in $commit'$ or $commit''$ to detect
such a cheater.  This concludes the proof.

%However, it is also useful to express the result in other terms.  
%In \cite{mayers95b}, our result 
%is translated in a language that do not involve quantum mechanics.
%More precisely, Bob's information about the bit $b$ is measured in
%terms of the probability $p_E$ that the best guess that Bob can
%make about the bit $b$ turns out to be wrong.  One must also take into
%account the probability $p_H$ that the bit $b$ is accepted by Bob when
%Alice is honest.  Normally, $p_H$ should be $1$ or very close to $1$
%and $p_E$ should be $\frac{1}{2}$ or very close to $\frac{1}{2}$.
%A lower bound on the probability $p_C$ with which Alice can
%successfully change her mind must be given in terms of $p_E$ and
%$p_H$.  The result of \cite{mayers95b} is $p_C > p_H - 4\sqrt{|p_E
%-\frac{1}{2}|}$.

\paragraph{Conclusions.}
Because we have shown that bit commitment is impossible, we cannot
hope to realize cryptographic 
primitives or applications which are known to be powerful enough to obtain
bit commitment.  On the other hand, there might exist
secure protocols for coin tossing and most multi-party
computations~\cite{kilian88,cgt95} because it is not known how
to build bit commitment on top of them.  
Note that some tasks might not be powerful enough to obtain
bit commitment and yet be impossible.  
What are the
fundamental principles that make some tasks possible and other tasks
impossible?  One could propose that all the tasks which involve only two
parties are impossible to explain why quantum key
distribution is possible and bit commitment impossible.  However,
there might be other principles involved.  For instance, in bit
commitment an asymmetry is created.  It could be that only the
asymmetrical tasks are impossible.  In this case, coin tossing would
be possible.  
What tasks are possible is a fundamental question which
yet remains to be answered.

\paragraph{Acknowledgments}  The author acknowledges fruitful discussions with
Charles Bennett,
Gilles Brassard, 
Claude Cr\'epeau, 
Lior Goldenberg, 
Jeroen van de Graaf,
Tal Mor, 
Louis Salvail,
Lev Vaidman, 
and William Wootters. 
The author also offers special thanks to the people of Maharishi University
of Management who provided a great support for the writing
of this letter. This work has been supported in part by 
DIMACS and by Qu\'ebec's FCAR.  

\end{document}